\documentclass[aps,prl,twocolumn,superscriptaddress,showpacs]{revtex4-1}

\usepackage{graphicx}
\usepackage[justification=raggedright]{caption}
\usepackage{amsfonts, amssymb, amsmath}
\usepackage{subcaption}
\newcommand{\circled}[1]{\raisebox{.5pt}{\textcircled{\raisebox{0pt} {{\scriptsize #1}}}}}
\newcommand{\MSbar}{\overline{\text{MS}}}
\newcommand\riken{RIKEN-BNL Research Center, Brookhaven National
  Laboratory, Upton, NY 11973, USA}
\newcommand\bnl{Brookhaven National Laboratory, Upton, NY 11973, USA}
\newcommand\edinb{SUPA, School of Physics, The University of
  Edinburgh, Edinburgh EH9 3JZ, UK}
\newcommand\cu{Physics Department, Columbia University, New York,
  NY 10027, USA}

\newcommand\soton{School of Physics and Astronomy, University of
  Southampton,  Southampton SO17 1BJ, UK}

\newcommand{\trinity}{School of Mathematics, Trinity College, 
Dublin 2, Ireland}

\begin{document}

%\preprint{}

\title{Emerging understanding of the $\mathbf{\Delta I = 1/2}$ Rule from Lattice QCD}


\author{P.A.~Boyle}\affiliation{\edinb}
\author{N.H.~Christ}\affiliation{\cu}
\author{N. Garron}\affiliation{\trinity}
\author{E.J. Goode}\affiliation{\soton} 
\author{T. Janowski}\affiliation{\soton} 
\author{C.~Lehner}\affiliation{\riken}
\author{Q.~Liu}\affiliation{\cu}
\author{A.T. Lytle}\affiliation{\soton} 
\author{C.T. Sachrajda}\affiliation{\soton}
\author{A.~Soni}\affiliation{\bnl}
\author{D.~Zhang}\affiliation{\cu}
%\thanks{}
\collaboration{The RBC and UKQCD Collaborations}


%\date{\today}

\begin{abstract}
There has been much speculation as to the origin of the  $\Delta I = 1/2$ rule (Re$A_0$/Re$A_2\simeq 22.5$). We find that the two dominant contributions to the $\Delta I=3/2$, $K\to\pi\pi$ correlation functions have opposite signs leading to a significant cancellation. This partial cancellation occurs in our computation of Re$A_2$ with physical quark masses and kinematics (where we reproduce the experimental value of $A_2$) and also for heavier pions at threshold. For Re$A_0$, although we do not have results at physical kinematics, we do have results for pions at zero-momentum with $m_\pi\simeq 420$\,MeV (Re$A_0$/Re$A_2=9.1(2.1)$) and $m_\pi\simeq 330$\,MeV (Re$A_0$/Re$A_2=12.0(1.7)$). The contributions which partially cancel in Re$A_2$ are also the largest ones in Re$A_0$, but now they have the same sign and so enhance this amplitude. The emerging explanation of the $\Delta I=1/2$ rule is a combination of the perturbative running to scales of $O(2\,$GeV), a relative suppression of Re$A_2$ through the cancellation of the two dominant contributions and the corresponding enhancement of Re$A_0$. QCD and EWP penguin operators make only very small contributions at such scales. 

\end{abstract}


%\keywords{}

\maketitle


%\section{Introduction \label{sec: Introduction}}
%\subsection{}
%\subsubsection{}
\begin{center} \textbf{Introduction}\end{center}
The ``$\Delta I=1/2$ rule" remains one of the longest-standing puzzles in particle physics. 
It refers to the surprising feature that in $K\to\pi\pi$ decays the final state is about 450 times more likely to have 
total isospin $I$=0 than $I$=2.
In terms of the (predominantly real) $K\to\pi\pi$ amplitudes $A_0$ and $A_2$, where the suffix denotes $I$, this corresponds to Re$A_0/$Re$A_2\simeq 22.5$.  
Perturbative running from the electroweak scale to about 1.5\,--2\,GeV contributes a factor of approximately 2 to this ratio~\cite{Gaillard:1974nj, Altarelli:1974exa}; the remaining factor of about 10 should come from non-perturbative QCD or, just possibly, from new physics. Lattice QCD provides the opportunity for the non-perturbative evaluation of $A_0$ and $A_2$, although it is only very recently that such direct $K\to\pi\pi$ calculations have become feasible.  In this letter we summarise the emerging explanation of the $\Delta I=1/2$ rule from computations of 
$A_0$ and $A_2$ by the RBC-UKQCD collaboration. 

The first results 
from direct simulations of a kaon decaying into two pions were presented in~\cite{Blum:2011pu, Blum:2011ng, Blum:2012uk}. 
The determination of $A_0$, where the two pions have vacuum quantum numbers, is particularly  challenging and so far it has not been calculated with physical masses and momenta.
We are striving to overcome technical issues such as the efficient evaluation of disconnected diagrams and the projection of the physical state through the use of G-parity boundary conditions~\cite{Wiese:1991ku,Kim:2002np,Kim:2003xt, Kim:2009fe} in order to evaluate $A_0$ at physical kinematics in the near future. In the meantime we have evaluated $A_0$ and $A_2$ for pions with masses of approximately 420\,MeV~\cite{Blum:2011pu} and 330\,MeV~\cite{qithesis} at threshold, i.e. with the pions at rest. For these unphysical masses we do find a significant enhancement of the ratio Re$A_0/$Re$A_2$, albeit a smaller one than 22.5 (see the first two rows of Table \ref{tab:params}). While investigating the origin of this enhancement we found a surprising cancellation in the evaluation of Re$A_2$, which significantly increases the ratio Re$A_0/$Re$A_2$.  This suppression of Re$A_2$ is the main result presented here.

\begin{table*}[t]
\begin{tabular}{c | c c c c c c | c}
&$~a^{-1}\,[\text{GeV}]$&$m_\pi \,[\text{MeV}]$ & $m_K\,\text{[MeV]}$ & Re$A_2\,[10^{\textrm{-}8}\, \text{GeV}]$ & Re$A_0 \,[10^{\textrm{-}8}\, \text{GeV}]$ &$\frac{\mathrm{Re}A_0}{\mathrm{Re}A_2}$ & notes \\
\hline
$16^3$ {\bf Iwasaki} &1.73(3)& 422(7) & 878(15) &4.911(31) &45(10)&9.1(2.1) &\,threshold calculation\\
$24^3$ {\bf Iwasaki} & 1.73(3)& 329(6) & 662(11) &2.668(14) &32.1(4.6)&12.0(1.7) &\,threshold calculation \\
$32^3$ {\bf IDSDR} & 1.36(1)& 142.9(1.1) & 511.3(3.9) &1.38(5)(26)&-& - &\,physical kinematics \\
${\bf Experiment}$ &--& 135\,-\,140&494\,-\,498&1.479(4)&33.2(2)&22.45(6)& 
\end{tabular}
\caption{Summary of simulation parameters and results obtained on three DWF ensembles. The errors with the Iwasaki action are statistical only, the second error for Re$A_2$ at physical kinematics from the IDSDR simulation is systematic and is dominated by an estimated 15\%
discretization uncertainty as explained in~\cite{Blum:2012uk}.
\label{tab:params}} 
\end{table*}

We have also evaluated $A_2$ with physical masses and momenta, obtaining a result for Re$A_2$ which agrees with the physical value and determining Im$A_2$ for the first time~\cite{Blum:2011ng, Blum:2012uk} (see the third row of Table 1). 
In the evaluation of Re$A_2$ at physical kinematics there is a similar cancellation; indeed it is even more pronounced than at the unphysical masses in the first two rows of Tab.\,\ref{tab:params}.

In the next section we summarize the simulations we have performed, highlighting features of immediate relevance for the $\Delta I = 1/2$ rule and referring to earlier publications for other details. We then explain the partial cancellation of the two contributions to Re$A_2$, which contradicts na\"ive expectations from the factorization (vacuum insertion) hypothesis. We also show that these two contributions have the same sign in Re$A_0$. We conclude by explaining how these features combine to provide an emerging understanding of the $\Delta I=1/2$ rule. Of course a full quantitative explanation will require a calculation of Re$A_0$ at physical kinematics which is underway. 
\begin{center}\textbf{Calculation of the Decay Amplitudes}\end{center}
Our evidence is based on calculations from three 
Domain Wall Fermion (DWF) ensembles with 2+1 sea-quark flavours (see Tab.\,\ref{tab:params}).  
Papers~\cite{Blum:2011ng, Blum:2012uk} describe a complete calculation of $A_2$ 
on a $32^3$ spacial lattice using the IDSDR (Iwasaki\,+\,Dislocation Suppressing Determinant Ratio) gauge action~\cite{Blum:2012yc}
for (almost) physical pion and kaon
masses and realistic kinematics. 
The  ensemble was generated at a single lattice spacing $a$ ($a^{-1}\!\simeq\! 1.4\,$GeV) chosen so that the volume is sufficiently large to accommodate the propagation of physical pions. 
In~\cite{Blum:2011pu} 
a complete calculation of both $A_0$ and  $A_2$ was carried out with the Iwasaki gauge action at $a^{-1}\simeq 1.7\,$GeV for $m_\pi\simeq 422$\,MeV and $m_K\simeq$\,737,~878 and 1117\,MeV (here we present results for $m_K\!\simeq 878$\,MeV which corresponds to almost energy-conserving decays).
Although the calculation was performed at threshold, this was the
first time a signal for Re$A_0$ had been obtained in the direct evaluation of the 
$K \rightarrow \pi \pi$ matrix elements.
A similar threshold calculation was presented in~\cite{qithesis} on a larger volume ($24^3$) with $m_\pi$= 329\,MeV.
The increased time extent of this lattice suppresses ``around-the-world'' effects in which one of the pions from the sink propagates in the forward time direction, crossing the periodic boundary and reaching the weak operator with the kaon.  
The calculation also used two-pion sources in which the single-pion wall sources are separated in time by a small number of time slices $\delta$ (the results
presented here are for $\delta=4$). We find that this suppresses the (unphysical) vacuum contributions in the 
$I=0$ channel, significantly reducing the noise. In this way  Re$A_0$ was resolved using only 138 configurations, 
compared to 800 in~\cite{Blum:2011pu}. With the actions used here, lattice artefacts scale parametrically as $O(a^2)$, although at present we are not in a position to take the continuum limit.

The amplitudes $A_0$ and $A_2$ can be expressed in terms of the ``master formula''
\begin{eqnarray} 
A_{I} &=& F_{I} \frac{G_F}{\sqrt{2}} V_{ud} V^{*}_{us} \sum_{i=1}^{10} \sum_{j=1}^{7} 
\Bigl[\bigl( z_i(\mu) + \bigr.\Bigr.\nonumber\\ 
&&\hspace{0.1in}\Bigl.\bigl. \tau y_i(\mu) \bigr)Z_{ij}^{\text{lat} \--> \MSbar} 
M_{j}^{\Delta I, \text{lat}} \Bigr] \qquad(I=0,2)
\,.
\label{master}
\end{eqnarray}
$\tau = -V^*_{ts}V_{td}/V^*_{us}V_{ud}$ and the $V_{ij}$ are elements of the Cabibbo-Kobayashi-Maskawa matrix.
$M_{i}^{\Delta I, \text{lat}} \equiv \langle(\pi \pi)_{I}\,|\,Q^{\text{lat}}_{i}\,|\, K\rangle$ are the matrix elements calculated on the lattice. They are determined by fitting three-point correlation functions composed of a kaon source at $t=0$,
a two-pion sink at $t=\Delta$, and one of the operators $Q_{i}^{\text{lat}}$ in the weak Hamiltonian inserted
at all times $0<t<\Delta$.   
 We fit the correlation functions $C_{I,i}(\Delta, t)$,
 \begin{equation} \label{C_2}
 C_{I,i}(\Delta, t) \approx M_{i}^{\Delta I,\mathrm{lat}} N_{\pi \pi} N_{K} e^{-E_{(\pi \pi)_I} \Delta} e^{-(m_K - E_{(\pi \pi)_I}) t}
 \end{equation}
 for $0 \ll t \ll \Delta$,  using a one parameter exponential fit to determine 
 the matrix elements $M_{i}^{\Delta I, \mathrm{lat}}$.
 $E_{(\pi \pi)_I}$ is the energy of the two-pion channel with isospin $I$. 
 All these correlation functions can be expressed in terms of the 48 contractions 
 enumerated in Section IV of~\cite{Blum:2011pu} and labelled
 $\circled{1}$ through $\circled{48}$. 
 The contractions are functions of $\Delta$ and $t$, 
 but we leave this dependence implicit, writing for example 
 $C_{2,1}(\Delta, t) = i \sqrt{2/3}\{ \circled{1} + \circled{2} \}$.
 
The renormalization factors $Z_{ij}^{\text{lat} \to \MSbar}$ provide the connection between
the bare lattice operators and those renormalized in the $\MSbar$--NDR scheme 
at the scale $\mu$,
\begin{equation}
Q^{\MSbar}_{i}(\mu) = Z_{ij}^{\text{lat}\--> \MSbar}(\mu, a)\, Q^{'\text{lat}}_j (a)\, .
\label{Zs}
\end{equation}
The operators $Q_{i}$ on the left of~\eqref{Zs} correspond to the conventional 10-operator
``physical" basis, which is over-complete (see e.g.\,\cite{Blum:2001xb}).
When calculating the renormalization factors, it is convenient to work in an equivalent
``chiral" basis of 7 linearly independent operators $Q^{'}_j$ with definite $SU(3)_\text{L} \times SU(3)_\text{R}$ transformation properties (see eqs.(172)-(175) in \cite{Blum:2001xb}).
$z_i(\mu) +  \tau y_i(\mu)$ are Wilson coefficient functions. $F_I$ is the Lellouch-L\"uscher factor relating the finite-volume Euclidean-space matrix element to the physical decay amplitude~\cite{Lellouch:2000pv}.

\textbf{Evaluation of Re$\mathbf{A_2}$:} $A_2$ receives contributions from the Electroweak Penguin (EWP) operators $Q_7$ and $Q_8$ as well as a single operator $Q^{3/2}_{(27,1)}$\,,
\begin{equation}\label{eq:q271}
Q^{3/2}_{(27,1)}=(\bar{s}^id^i)_L\big\{(\bar{u}^ju^j)_L\!-\!(\bar{d}^jd^j)_L\big\}+(\bar{s}^iu^i)_L(\bar{u}^jd^j)_L, 
\end{equation}
where the superscript $3/2$ denotes $\Delta I$ and the subscript $(27,1)$ denotes how the operator transforms under $SU(3)_L\times SU(3)_R$ chiral symmetry. $i,j$ are color labels and the spinor indices are contracted within each pair of parentheses. The subscript $L$ denotes \emph{left}, so that e.g. 
$(\bar{s}^id^i)_L(\bar{u}^ju^j)_L=(\bar{s}^i\gamma^\mu(1-\gamma^5)d^i)\,(\bar{u}^j\gamma_\mu(1-\gamma^5)u^j)$.
The $\Delta I=3/2$ components of the operators $Q_1, Q_2, Q_9$ and $Q_{10}$ are all proportional to $Q^{3/2}_{(27,1)}$. From all our simulations we confirm that the contribution from the EWP operators to Re$A_2$ is about 1\%; e.g. for physical kinematics we find Re$A_2=(1.381\pm0.046\pm0.258)\,10^{-8}$\,GeV to which the EWP operators contribute $-0.0171\,10^{-8}$\,GeV~\cite{Blum:2011ng,Blum:2012uk} (the physical value is Re$A_2=1.479(4)10^{-8}$\,GeV). We therefore neglect the EWP operators in the following discussion. Chiral symmetry implies that $Q^{3/2}_{(27,1)}$ does not mix with the EWP operators so that Re$A_2$ is proportional to its lattice matrix element; the constant of proportionality is the product of the Wilson coefficient, the renormalization constant, finite-volume effects and kinematical factors (see \cite{Blum:2012uk} for a detailed discussion, including an explicit demonstration that the mixing is indeed negligible in the DWF simulation). 

\begin{figure}
\begin{subfigure}[h]{.238\textwidth}
\includegraphics[width=\textwidth]{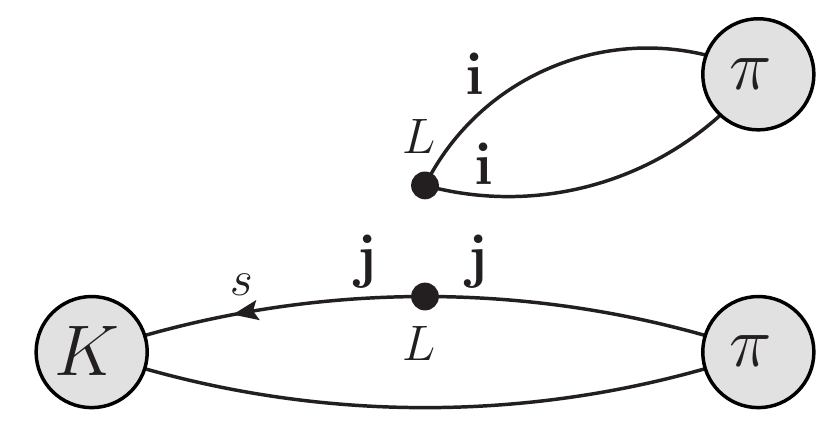}
\caption*{Contraction \protect\circled{1}.}
\end{subfigure}
\begin{subfigure}[h]{.238\textwidth}
\includegraphics[width=\textwidth]{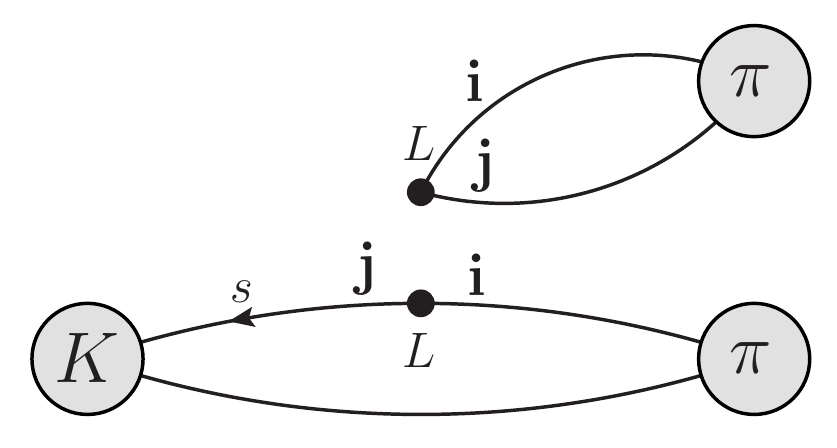}
\caption*{Contraction \protect\circled{2}.}
\end{subfigure}
\caption{The two contractions contributing to Re$A_2$. They are distinguished by the color summation ($\mathbf{i,j}$ denote color). $s$ denotes the strange quark and $L$ that the currents are  left-handed. \label{fig:c1c2}}
\end{figure}

\begin{figure}[t]
\includegraphics[width=.45\textwidth]{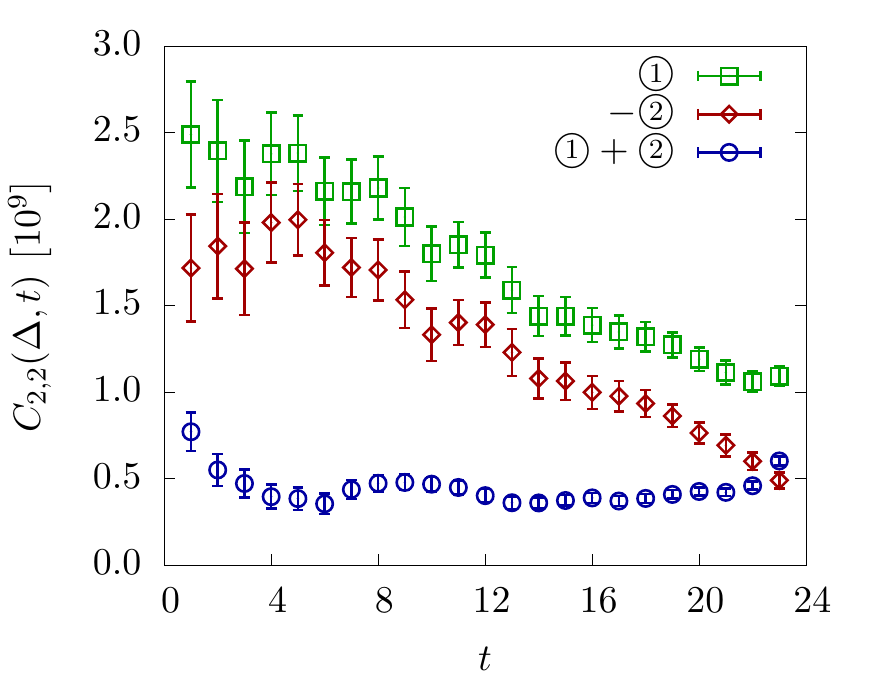}
\caption{Contractions $\protect\circled{1}$, -$\protect\circled{2}$ and $\protect\circled{1}+\protect\circled{2}$ as functions of $t$ from the simulation at physical kinematics and with $\Delta=24$ . \label{fig:c1c2physical}}
\end{figure}

\begin{figure}[t]
\includegraphics[width=.45\textwidth]{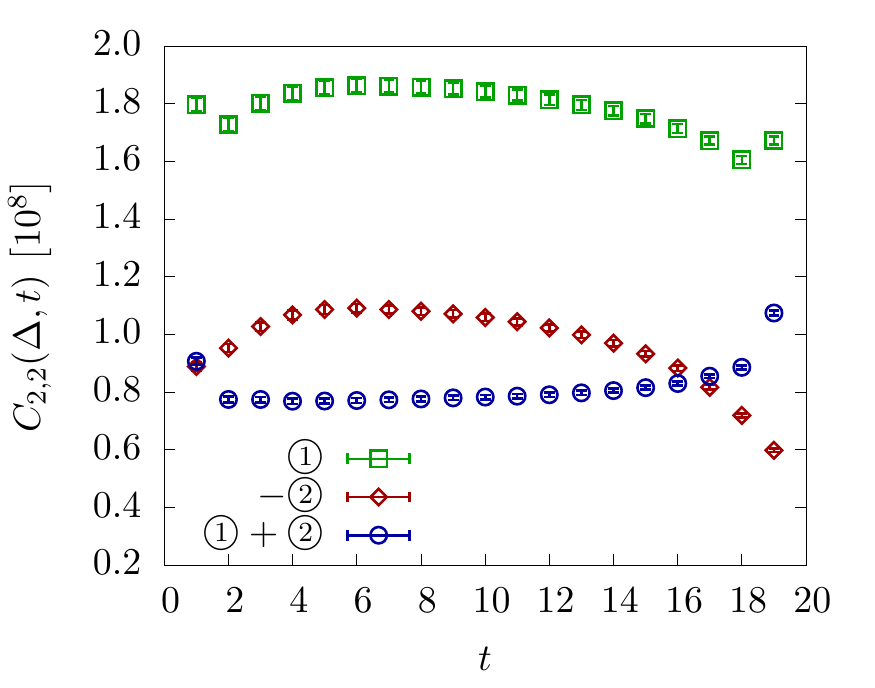}
\caption{Contractions $\protect\circled{1}$, -$\protect\circled{2}$ and $\protect\circled{1}+\protect\circled{2}$ as functions of $t$ from the simulation at threshold with $m_\pi\simeq$ 330\,MeV and $\Delta=20$. \label{fig:c1c2330}}
\end{figure}

Fierz transformations allow the $K\to\pi\pi$ correlation function of $Q^{3/2}_{(27,1)}$ to be reduced  to the sum of the two contractions illustrated in Fig.\,\ref{fig:c1c2}, labeled by $\circled{1}$ and $\circled{2}$. The two contractions are identical except for the way that the color indices are summed. $A_2$ is proportional to the matrix element extracted from the sum $\circled{1}+\circled{2}$. The main message of this letter is our observation from all three simulations that $\circled{1}$ and $\circled{2}$ have opposite signs and are comparable in size. This is illustrated in Fig.\,\ref{fig:c1c2physical} for the results at physical kinematics from~\cite{Blum:2011ng,Blum:2012uk}, where we plot $\protect\circled{1}$, -$\protect\circled{2}$ and $\protect\circled{1}+\protect\circled{2}$ as functions of $t$.  We extract $A_2$ by fitting $\circled{1}+\circled{2}$ in the interval $t\in[5,19]$ where there is a significant cancellation between the two terms. A similar, although not quite so pronounced cancellation occurs at threshold for physical masses  and for the heavier masses studied in \cite{Blum:2011pu,qithesis}, see Fig.\,\ref{fig:c1c2330} for example. 

We stress that it is only the correlation function $\circled{1}+\circled{2}$ which has a time behaviour corresponding to $E_{(\pi\pi)_{2}}$. Because the calculation is performed in a finite-volume $E_{(\pi\pi)_2}\neq E_{(\pi\pi)_0}$ and $\circled{1}$ and $\circled{2}$ individually have an isospin\,0 component. If $E_{(\pi\pi)_{2}}=m_K$ then $\circled{1}+\circled{2}$ is independent of $t$ away from the kaon and two-pion sources, and this is what we observe, particularly in Fig.\,\ref{fig:c1c2physical} where the energies are matched most precisely.

It has been argued that the factorisation hypothesis\,\cite{Gaillard:1974hs} works reasonably well in reproducing the experimental value of $A_2$ (see e.g. Sec.VIII-4 in \cite{Donoghue:1992dd}). In this approach, the gluonic interactions between the quarks combining into different pions are neglected and $A_2$ is related to the decay constant $f_{\pi}$ and the $K_{\ell 3}$ form factor close to zero momentum transfer. On the basis of color counting, one might therefore expect that  $\circled{2}\simeq 1/3\,\circled{1}$, whereas, for physical kinematics, we find $\circled{2}\simeq - 0.7\circled{1}$ and that nevertheless $\circled{1}+\circled{2}$ leads to the correct result for $A_2$. Thus the expectation based on the factorisation hypothesis proves to be unreliable here.

Following the discovery that $\circled{1}$ and $\circled{2}$ have opposite signs we examined separately the two contributions to the matrix element $\langle \bar{K}^0|(\bar{s}d)_L
(\bar{s}d)_L|K^0\rangle$ which contains the non-perturbative QCD effects in neutral kaon mixing~\cite{Blum:2012yc}. The two contributions correspond to Wick contractions in which the two quark fields in the $K^0$ interpolating operator are contracted i) with fields from the same current in $(\bar{s}d)_L
(\bar{s}d)_L$ and ii)~with one field from each of the two currents. Color counting and the vacuum insertion hypothesis suggest that the two contributions come in the ratio 1:1/3, whereas we find that in QCD they have the opposite sign.  This had been noticed earlier; see e.g.~\cite{Lellouch:2011qw} and references therein. 

We postpone a discussion of the implications of these results to the $\Delta I=1/2$ rule until the next section, but we believe that the partial cancellation observed in the evaluation of $A_2$ is a significant component.

\textbf{Evaluation of Re$\mathbf{A_0}$:}
The evaluation of $A_0$ at physical kinematics has not yet been completed. The results  presented here are obtained at threshold, with the two pions in their zero-momentum ground state with each pion at rest up to finite-volume effects. Even at threshold we have had to
overcome many theoretical and technical problems, including the evaluation of the 48 contractions  contributing to the correlation functions, the renormalization of the operators in the effective Hamiltonian, the subtraction of power divergences and the evaluation of the finite-volume corrections. The threshold calculations do not require however, the isolation of an excited state. The pions in a physical decay each have a non-zero momentum in the center-of-mass frame, which corresponds to an excited state in lattice calculations. Given the poor statistical signals after the subtraction of power divergences and the evaluation of disconnected diagrams, the evaluation of $A_0$ at physical kinematics is currently impracticable with standard techniques and is the main motivation for our development of G-parity boundary conditions\,~\cite{Wiese:1991ku, Kim:2002np, Kim:2003xt, Kim:2009fe}.    

With the two pions at threshold we find\,\cite{Blum:2011pu,qithesis}
\begin{equation}\label{eq:ratios}
\frac{\mathrm{Re}A_0}{\mathrm{Re}A_2}=\begin{cases} 
	9.1(2.1)~~\mathrm{for~}m_K=878\,\mathrm{MeV},\,m_\pi=422\,\mathrm{MeV}
 \\ 12.0(1.7)\,\mathrm{for~}m_K=662\,\mathrm{MeV},\,m_\pi=329\,\mathrm{MeV}.
\end{cases}
\end{equation}
While these results differ significantly from the observed value of 22.5, because the calculations are not performed at physical kinematics, there is nevertheless already a significant enhancement in the ratio and it is interesting to understand its origin. In Tab.\,\ref{tab:a0breakdown} we present the contributions to Re$A_0$ from each of the lattice operators in the $24^3$ simulation with $a^{-1}=1.73(3)$\,GeV and from each $\MSbar$-$\mathrm{NDR}$ operator at renormalization scale 2.15\,GeV. In both cases, the dominant contribution comes from the current-current operators $Q_2$. 

Since in a finite-volume $E_{(\pi\pi)_2}\neq E_{(\pi\pi)_0}$,
one cannot satisfy the condition $m_K=E_{\pi\pi}$ for both isospin channels simultaneously with the same quark masses. Here we quote results using the fixed meson masses quoted in Eq.\,(\ref{eq:ratios}), which is sufficient for our current discussion. For these masses $E_{(\pi\pi)_0}=766(29)$\,MeV (629(15)\,MeV), $E_{(\pi\pi)_2}=876(15)$\,MeV (668(11)\,MeV) for the $16^3$ ($24^3$) lattice.
A study that interpolates in the kaon mass to make both decays energy-conserving may be found in~\cite{Blum:2011pu}.

\begin{table}
\begin{tabular}{c|c|c}
i & $Q_i^{\text{lat}}\;$[GeV]&$Q_i^{\MSbar \text{-NDR}} \;$[GeV]\\ \hline
1&\,\phantom{-}8.1(4.6) $10^{-8}\,\,$&\phantom{-}6.6(3.1) $10^{-8}\,\,$\\ 
2&\,\phantom{-}2.5(0.6) $10^{-7}\,\,$&\phantom{-}2.6(0.5) $10^{-7}\,\,$\\ 
3&\,-0.6(1.0) $10^{-8}\,\,$&\phantom{-}5.4(6.7) $10^{-10}$\\ 
4&--&\phantom{-}2.3(2.1)~$10^{-9}\,\,$\\ 
5&\,-1.2(0.5) $10^{-9}\,\,$&\phantom{-}4.0(2.6) $10^{-10}$\\ 
6&\,\phantom{-}4.7(1.7) $10^{-9}\,\,$&-7.0(2.4) $10^{-9}\,\,$\\ 
7&\,\phantom{-}1.5(0.1) $10^{-10}$&\phantom{-}6.3(0.5) $10^{-11}$\\ 
8& -4.7(0.2) $10^{-10}$&-3.9(0.1) $10^{-10}$\\ 
9&--&\phantom{-}2.0(0.6) $10^{-14}$\\ 
10& --& \phantom{-}1.6(0.5) $10^{-11}$\\ \hline
Re$A_0$ &\,\phantom{-}3.2(0.5) $10^{-7}\,\,$ & \phantom{-}3.2(0.5) $10^{-7}\,\,$
\end{tabular}
\caption{\label{tab:a0breakdown} 
Contributions from each operator to Re$A_0$ for $m_K=662$\,MeV and $m_\pi=329$\,MeV. The second column contains the contributions from the 7 linearly independent lattice operators with $1/a=1.73(3)$\,GeV and the third column those in the 10-operator basis in the $\MSbar$-$\mathrm{NDR}$ scheme at $\mu=2.15$\,GeV. Numbers in parentheses represent the statistical errors.}
\end{table}

\begin{figure}[t]
\includegraphics[width=0.45\textwidth]{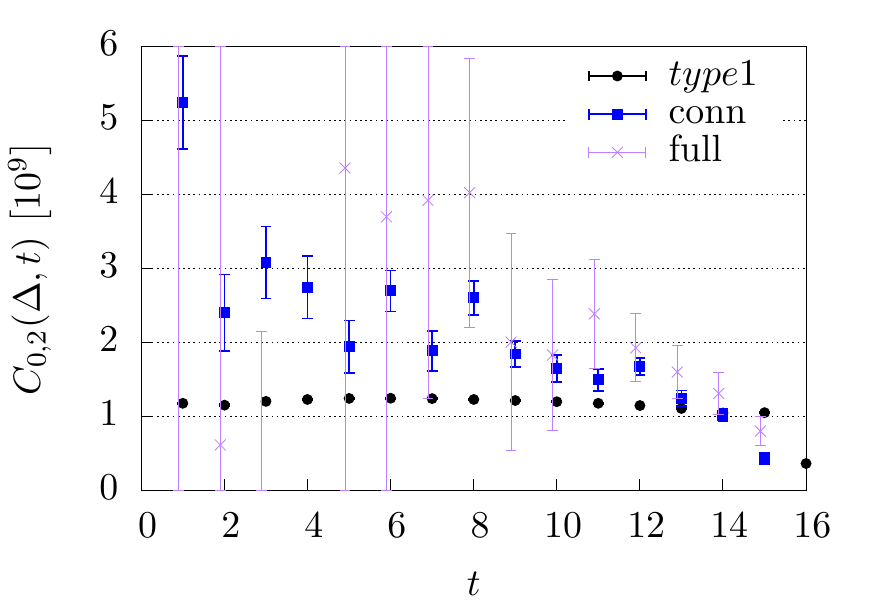}
\caption{Contributions of $Q_{2}^{\text{lat}}$ to Re$A_0$ (purple crosses).  
The blue squares and black circles denote the connected and \emph{type1} contractions respectively.
\label{fig:I0_Q2}}
\end{figure}

The dominant contribution from the lattice operator $Q_2$ to the $\Delta I=1/2$ correlation function is proportional to the contractions $2 {\cdot} \circled{1}-\circled{2}$ and corresponds to \emph{type1} diagrams in the language of~\cite{Blum:2011pu} (see Fig.3 in~\cite{Blum:2011pu}). In Fig.~\ref{fig:I0_Q2} we show the total contribution of $Q_2$ to the correlation function, as well as the total connected contribution and that of \emph{type1} diagrams given by
$\frac{i}{\sqrt{3}}\{2 {\cdot} \circled{1} - \circled{2}\}$. The errors on the total contribution are dominated by the disconnected diagrams. The observation that $\circled{1}$ and $\circled{2}$ have opposite signs leads to an enhancement between the two terms rather than the suppression in the factorization 
approximation $\circled{2}=\frac{1}{3}\circled{1}$.
Similarly, in the case of $Q_1$, the \emph{type1} combination 
$\frac{i}{\sqrt{3}}\{2 {\cdot} \circled{2} - \circled{1}\}$ is dominant.
In this case both the correlation function and the Wilson coefficient 
$z_{1}(\mu) + \tau y_{1}(\mu)$ are negative, so that the overall contribution adds to that from the correlation function of $Q_2$.

Finally we note that in our data Re$A_2$ shows a much stronger mass dependence than Re$A_0$, which was also expected in SU(2) chiral perturbation theory~\cite{Bijnens:2009yr}. We attribute this to the partial cancellation between $\circled{1}$ and $\circled{2}$ in Re$A_2$. Our results for Re$A_2$ and Re$A_0$ are given in Tab.\ref{tab:params}.

\begin{center}\textbf{Conclusions}\end{center} From our recent computations of $K\to\pi\pi$ decay amplitudes a likely explanation of the $\Delta I=1/2$ rule is emerging. In particular, we find that in the evaluation of Re$A_2$, which is proportional to the sum of two contractions $\circled{1}+\circled{2}$, there is a significant cancellation between the two terms.
The na\"ive expectation based on the factorization hypothesis suggests that $\circled{2}\approx \frac13\circled{1}$, whereas in QCD we find that they have the opposite sign. (The two terms contributing to $B_K$ similarly have opposite signs contradicting expectations from the vacuum insertion approximation.) 

The evaluation of $A_0$ at physical kinematics has not yet been performed. Our simulations at threshold with $m_\pi=329$\,MeV and 422\,MeV show that the dominant contributions to $A_0$ comes from the current-current operators, with only small corrections from the penguin operators. This is true whether we express the results in terms of the bare lattice operators at $a^{-1}=1.73\,$GeV or the $\MSbar$-NDR renormalized operators at $\mu=2.15$\,GeV (see Tab.\,\ref{tab:a0breakdown}). Although 48 contractions contribute to the $I=0$ correlation function, in our simulations the largest contributions again come from contractions $\circled{1}$ and $\circled{2}$ with relative signs which enhance Re$A_0$. 

References to estimates of the amplitudes using analytic or model approximations are presented in the reviews~\cite{Buchalla:1995vs,Cirigliano:2011ny}. We note that a suppression of Re$A_2$ and an enhancement of Re$A_0$ was found in~\cite{Bardeen:1986vz} using the $1/N$ expansion with a particular ansatz for matching the short and long-distance factors at scales 0.6-0.8\,GeV.  

The results presented above indicate that Re$A_2$ is very sensitive to the choice of quark masses and momenta; a sensitivity we attribute to the partial cancellation of the two contributing contractions. On the other hand, there is no such cancellation in Re$A_0$ and indeed the results depend much less on the masses and the values we find are already close to the experimental result.
Of course before we can claim to understand the $\Delta I=1/2$ rule quantitatively, we need to reproduce Re$A_0$/Re$A_2$=22.5 at physical quark masses and kinematics and in the continuum limit and we are currently undertaking this challenge. Nevertheless, from the results and discussion of this paper it appears that, in addition to the well known perturbative enhancement of Re$A_0$/Re$A_2$, the explanation is a combination of a significant relative suppression of Re$A_2$ as well as some enhancement of Re$A_0$ with penguin operators contributing very little. 
\\

\textbf{Acknowledgements} We thank W.\,Bardeen and A.\,Buras for informative discussions. P.\,Boyle was supported in part by STFC grants ST/J000329/1, ST/K005804/1,
ST/K000411/1 and ST/H008845/1, 
N.\,Christ, Q.\,Liu and D.\,Zhang by U.S.~DOE grant DE-FG02-92ER40699, E.\,Goode, T.\,Janowski, A.\,Lytle and C.\,Sachrajda by STFC Grant ST/G000557/1, C.\,Lehner by the RIKEN FPR program and A.\,Soni by US DOE grant DE-AC02-98CH10886(BNL).
\bibliography{refs}

\end{document}